\documentclass[preprint,aps,showpacs]{revtex4}       
\usepackage{graphicx}
\newcommand{\beq}{\begin{equation}}
\newcommand{\eeq}{\end{equation}}
\begin{document}
\title{Hindrance of $^{16}$O+$^{208}$Pb fusion at extreme sub-barrier energies}
\author{Henning Esbensen}
\affiliation{Physics Division, Argonne National Laboratory, Argonne, Illinois 60439}
\author{\c{S}erban Mi\c{s}icu}
\affiliation{National Institute for Nuclear Physics, Bucharest, P. O. Box MG6, Romania}
\date{\today}
\def\o16{$^{16}$O}
\def\pb208{$^{208}$Pb}

\begin{abstract}
We analyze the fusion data for $^{16}$O+$^{208}$Pb using coupled-channels 
calculations. We include couplings to the low-lying surface excitations 
of the projectile and target and study the effect of the ($^{16}$O,$^{17}$O) 
one-neutron pickup. 
The hindrance of the fusion data that is observed at energies far below 
the Coulomb barrier cannot be explained by a conventional ion-ion potential 
and defining the fusion in terms of ingoing-wave boundary conditions (IWBC).
We show that the hindrance can be explained fairly well by applying the M3Y 
double-folding potential which has been corrected with a calibrated, 
repulsive term that simulates the effect of nuclear incompressibility.

We show that the coupling to one-neutron transfer channels plays a crucial 
role in improving the fit to the data.  
The best fit is achieved by increasing the transfer strength by 25\%
relative to the strength that is required to reproduce the one-neutron 
transfer data.  The larger strength is not unrealistic because the 
calculated inelastic plus transfer cross section is in good agreement 
with the measured quasielastic cross section.
We finally discuss the problem of reproducing the fusion data at energies far 
above the Coulomb barrier.  Here we do not account for the data when we apply 
the IWBC but the discrepancy is essentially eliminated by applying the 
M3Y+repulsion potential and a weak, short-ranged imaginary potential.
\end{abstract}
\maketitle

\section{Introduction}

It would be desirable to be able to make a consistent coupled-channels 
calculation of the most important reaction channels in \o16 on \pb208 
collisions at energies near and below the Coulomb barrier.  
This has been attempted several times in the past \cite{Steve,Ian1,Ian2} 
but the analyses were never completely satisfactory. 
For example, it was difficult to reproduce the energy 
dependence of the measured fusion cross section \cite{Flem,Vulg}. 
In order to improve the fit to the data it was necessary either to use a 
very small diffuseness of the ion-ion potential \cite{Steve} or to use a 
complex ion-ion potential in the calculations \cite{Ian1,Ian2}.
Both modifications indicate that the calculations were either incomplete 
in terms of the reaction channels that were considered or that 
other features of the calculations were unrealistic.

The old $^{16}$O+$^{208}$Pb fusion data \cite{Flem,Vulg} turned out to 
contain some uncertainties, in particular with respect to the evaporation 
residue component (see Ref. \cite{Morton1}). 
The most accurate data that are now available can be found in Ref. 
\cite{Morton2}.
The analysis of the revised data showed that there are still some
inconsistencies with coupled-channels calculations.
Thus it was necessary to use a large diffuseness of the ion-ion potential
in order to reproduce the fusion data at energies above the Coulomb barrier, 
whereas the fusion barrier distribution extracted from the data required 
a very small diffuseness.

We have recently pointed out \cite{Misicu1} that many of the 
ion-ion potentials, which have been used in the past, are unrealistic 
at small distances between the reacting nuclei. While the ion-ion potential
apart for minor adjustments is quite accurately given by the M3Y 
double-folding potential at larger distances (say, outside the 
Coulomb barrier), this potential is unrealistic at smaller distances, 
where it produces a pocket in the entrance channel potential that is far 
too deep, sometimes even deeper than the ground state energy of the 
compound nucleus.
By considering the effect of nuclear incompressibility we obtained what we
think is a more realistic interaction, which we call the M3Y+repulsion 
potential. It produces a rather shallow pocket in the entrance channel  
potential \cite{Misicu1}. Such a shallow pocket 
% is a prerequisite for 
makes it possible to accurately reproduce 
the measured fusion cross sections for 
$^{64}$Ni+$^{64}$Ni \cite{Jiangni}, in particular at the lowest energies, 
where the data fall off steeply with decreasing energy. 
The steep falloff is referred to as the fusion hindrance; see Ref. 
\cite{Jiangsys} for a recent discussion of this phenomenon.

The measurements of the fusion of $^{16}$O+$^{208}$Pb \cite{Morton2} 
were recently extended 
to very small cross sections \cite{Morton3}. The new data exhibit a fusion
hindrance at low energies which is similar to what has been observed for many 
other heavy-ion systems \cite{Jiangsys}. 
It is therefore of interest to see whether the M3Y potential, corrected for 
nuclear incompressibility as discussed above, can account for the new data
when applied in coupled-channels calculations.
It is also of interest to investigate whether the
M3Y+repulsion potential can explain the suppression of the 
high-energy fusion data which was discussed in Ref. \cite{Newton}.

The coupled-channels calculations that have been performed previously 
\cite{Steve,Ian1,Ian2,Morton2} included couplings to the $2^+$, $3^-$,
and $5^-$ low-lying states in $^{208}$Pb, the lowest $3^-$ state in \o16,
and to transfer channels [(\o16,$^{17}$O) neutron pickup and 
(\o16,$^{15}$N) proton stripping]. The reaction data 
\cite{Flem} had a significant yield of C isotopes, which were simulated 
by simplified couplings in some of the the calculations \cite{Steve,Ian2}.
The calculations showed that the couplings to the transfer channels have
a significant effect on the calculated fusion cross sections and improve
the fit to the data.
% It was noticed in Ref. \cite{Gupta} that couplings to the two-phonon octupole 
% state in Pb improved the fit to the barrier distribution extracted from the 
% fusion data but the quality of the fit to the actual fusion data did not 
% appear to improve much.

We include in our analysis of the $^{16}$O+$^{208}$Pb fusion data 
\cite{Morton2,Morton3} 
some of the most important surface excitation modes and study the effect 
of the (\o16,$^{17}$O) neutron pickup reaction, which is one of 
the most dominant reaction channels besides fusion \cite{Flem,Pieper}.
We also study how well we can account for the total reaction cross
section and the elastic scattering data \cite{Flem}.

\section{Description of the calculations}

The coupled-channels calculations we perform are similar to those discussed 
in Refs. \cite{esblan89,nisn}. Here we summarize the approximations we make
and describe the input to the calculations.  
The basic assumption is the rotating frame approximation, which implies that 
one has to include only one channel for each state of spin $I$, and not the 
$I+1$ (or even $2I+1$) channels that are required in general.
This approximation is commonly used in calculations of heavy-ion fusion cross 
sections because it makes it possible to include the effect of many reaction 
channels.
It is a reliable approximation for calculating fusion and elastic scattering
but the angular distributions for inelastic scattering and transfer reactions
can be poor, in particular at forward angles 
(see, for example, Ref. \cite{esblan89}).  

\subsection{Ion-ion potentials} 

The ion-ion potentials we use are the same as those we applied in Ref. 
\cite{Misicu2}, namely, the Ak\"uyz-Winther (AW) and the M3Y+repulsion 
double-folding potential.
The AW potential is defined by Eqs. (40,41,44,45) in Sect. III.1 of Ref. 
\cite{BW} but we have modified Eq. (40) by introducing an adjustable 
radius parameter $\Delta R$,
\beq
U_{12}^N(r) = \frac{- 16\pi\gamma R_{12}a}{1+\exp((r-R_1-R_2-\Delta R)/a)}.
\label{AWpot}
\eeq
Here $\gamma$ = 0.95 MeV/fm$^2$ is the nuclear surface tension, 
$a$ is the surface diffuseness defined in Eq. (44), Sect. III.1 of 
Ref. \cite{BW}, $R_i=1.2A_i^{1/3}-0.09$ fm, and $R_{12}=R_1R_2/(R_1+R_2)$. 
The parameter $\Delta R$ is adjusted so that the two potentials,
namely, the AW and the M3Y+repulsion potentials, produce the same 
Coulomb barrier height.
In the case discussed below this requires the value $\Delta R$ = 0.13 fm.

The M3Y double-folding potential is calculated numerically 
in the Fourier representation, 
\beq
U(r) = \frac{1}{2\pi^2} \int q^2dq \
\rho_1(q) \ \rho_2(q) \ v(q) \ j_0(qr).
\label{dfp}
\eeq
Here $v(q)$ represents the M3Y effective nucleon-nucleon interaction,
$j_0(x)$ is the spherical Bessel function, and $\rho_{i}(q)$ is the 
Fourier transform of the density of nucleus $i$.
The Yukawa and contact type interactions that we use \cite{Misicu2} to 
define the direct and exchange part of the M3Y interaction have simple 
analytic Fourier transforms.

The densities are parametrized in terms of a Fermi function.
We use the proton and neutron density parameters: $R_p$ = 2.53, $R_n$ = 2.57, 
$a$=0.513 fm for $^{16}$O, and $R_p$ = 6.60, $R_n$ = 6.75, $a$=0.546 fm for 
$^{206}$Pb. The proton densities are consistent with the measured charge 
densities \cite{DeVries} while the radii for the neutron densities are 
slightly larger. The Fermi function does not have a simple analytic Fourier 
transform so we use instead an accurate analytic approximation, which has 
an exact analytic Fourier transform. 
This approximation is described in the appendix.

The repulsive interaction, which simulates the effect of the nuclear 
incompressibility (see Ref. \cite{Misicu2} for details) is calculated 
from the same integral, Eq. (\ref{dfp}). In this case the $v(q)$  
represents the (constant) Fourier transform of a contact interaction 
with the strength $V_{rep}$. The densities we use in the integral have
the same radii as those used in the calculation of the M3Y double-folding 
potential but the diffuseness $a_{rep}$ is chosen differently.
We have chosen the parameters $V_{rep}$ = 570 MeV~fm$^3$ and $a_{rep}$ = 
0.35 fm and obtain a nuclear incompressibility of 245 MeV.  
The total potential (M3Y+repulsion plus Coulomb interaction) is 
illustrated by the solid curve in Fig. 1.  It has a Coulomb barrier of 
75.6 MeV and the minimum energy of the pocket is 65.1 MeV. The latter 
pocket energy was chosen because it is required by the fusion data as 
we shall see later on, and this was achieved by adjusting $a_{rep}$.

\begin{figure}
\includegraphics[width = 8cm]{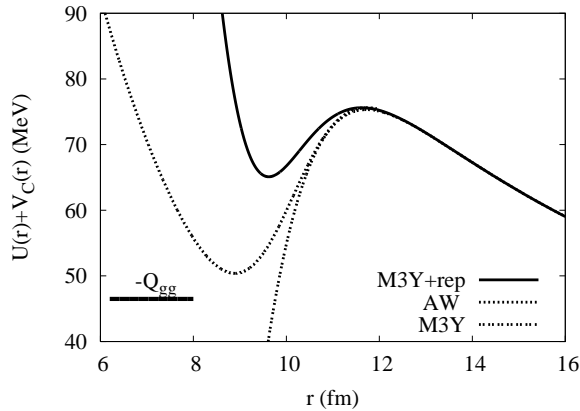}
\caption{\label{pot} The M3Y and M3Y+repulsion potentials are compared 
to the AW potential. The height of the Coulomb 
barrier is 75.6 MeV in all three cases.}
\end{figure}

The AW potential, Eq. (\ref{AWpot}), with $\Delta R$ = 0.13 fm adjusted 
to produce the same Coulomb barrier height as the M3Y+repulsion potential, 
is also shown in Fig. \ref{pot}. It has a pocket energy at 50.5 MeV.
The total potential which is based on the pure (direct+exchange) M3Y 
potential has a much deeper pocket. It is unrealistic because the minimum 
is lower than the ground state energy of the compound nucleus $^{224}$Th, 
which is indicated in the figure by the thick solid, horizontal line 
at -$Q_{gg}$ = 46.5 MeV. 

\subsection{Surface excitations}

The structure input that will be used to describe the excitation of the
low-lying states in $^{16}$O and $^{208}$Pb is given in Table I. 
For the lead states
it is assumed that the Coulomb and nuclear $\beta$-values are identical.
For the oxygen states we use the parameters that were determined in 
Ref. \cite{Knopf}.  The coupling strengths to the oxygen states are rather 
strong, in particular for the octupole state, but the strengths for Coulomb 
excitation are consistent with the adopted values \cite{ENDF}.  
In addition to the states shown in Table I we also include all mutual 
excitations of these states up to an excitation energy cutoff of 7.7 MeV. 
That gives a total of 12 channels. 

\begin{table}
\caption{Structure input for $^{16}$O \cite{Knopf,ENDF} and $^{208}$Pb 
\cite{ENDF} The quadrupole moments of the $2^+$ and $3^-$ states in \pb208,
$Q$ = -0.7(3) and -0.34(15) b \cite{Ragh},  respectively, have been 
converted into an effective quadrupole deformation $\beta(Q)$.}
\begin{tabular} {|c|c|c|c|c|c|c|}
\colrule
Nucleus &
$\lambda^\pi$ & E$_x$ (MeV) & 
B(E$\lambda$) (W.u.) & $\beta_\lambda^C$ & 
$\beta_\lambda^N$ & $\beta(Q)$ \\
\colrule
$^{16}$O   & $2^+$ & 6.92  &  3.1(1)  & 0.352 & 0.324 &  -   \\
           & $3^-$ & 6.13  & 13.5(7)  & 0.713 & 0.481 &  -   \\
\colrule
$^{208}$Pb & $3^-$ & 2.615 & 33.9(5)  & 0.111 & 0.111 & 0.038 \\
           & $5^-$ & 3.198 & 11.0(7)  & 0.059 & 0.059 &  -   \\
           & $2^+$ & 4.085 &  8.7(5)  & 0.057 & 0.057 & 0.078 \\
           & $4^+$ & 4.323 & 18.0(13) & 0.079 & 0.079 &  -   \\
\colrule
\end{tabular}
\end{table}

As in our previous work \cite{Misicu2} we include all couplings up to 
second order in the nuclear deformation parameters, whereas Coulomb 
excitation is described by linear couplings. The form factors for the 
linear and quadratic nuclear couplings are assumed to be the first 
and second radial derivatives of the ion-ion potential, respectively.

\subsection{Neutron transfer}

We will also study the effect of transfer and consider explicitly the 
one-neutron transfer to the 5/2$^+$ ground state of $^{17}$O leaving 
the $^{207}$Pb nucleus in the $1/2^-$ ground state, and in the $5/2^-$ 
and $3/2^-$ excited  states at 0.57 and 0.88 MeV, respectively. 
The Q-values for these transfers are -3.22 MeV, -3.79, and -4.10 MeV.
These transfer channels dominate the measured ($^{16}$O,$^{17}$O) cross 
section at 104 MeV lab energy \cite{Pieper}, and the spectroscopic 
factors that were extracted are close to one.
The three transfer channels are lumped together into one effective 
transfer channel, the same way it was done in Ref. \cite{esblan89}.
The effective Q-value for the transfer is set to $Q_{eff}$ = -3.2 MeV. 
This represents a weighted average of the actual Q-values
corrected as suggested in Ref. \cite{broglia} for the lower Coulomb 
barrier ($\Delta V_{CB}$ = 0.46 MeV) in the transfer channel, i.~e., 
\beq
Q_{eff} = 
\frac{\sum Q_{n} \sigma_{n}}{\sum \sigma_{n}} \
+ \ \Delta V_{CB}.
\eeq
Here the sums over $N$ are over the three final states mentioned above.

The transfer form factors we use are taken from Ref.  \cite{Quesada} 
and they are calculated for full spectroscopic strength.
The overall strength of the effective form factor will be scaled 
by the factor $F_{1n}$, in order to be able to reproduce the measured
transfer data \cite{Pieper} at 104 MeV in the laboratory frame. 
We shall see later on (Fig. \ref{reacom3y}B) that this requires the strength 
$F_{1n}$ = 0.95, whereas boosting the strength to $F_{1n}\approx$ 1.2
makes it possible to simulate the total reaction cross section.

\section{Results of the calculations}

The coupled-channels equations are solved with the usual scattering
conditions  at large distances and ingoing-wave boundary conditions 
(IWBC) that are imposed at the location $R_{\rm pocket}$ of minimum 
of the potential pocket in the entrance channel.
We found in Ref. \cite{Misicu2} that the fusion 
hindrance observed at extreme subbarrier energies could only be 
explained by defining the fusion in terms of IWBC.
However, in order to be able to simulate the fusion  and the total 
reaction cross section at energies far above the Coulomb barrier 
we will also consider the possibility of supplementing the IWBC with 
an imaginary potential.

\begin{figure}
\includegraphics[width = 9cm]{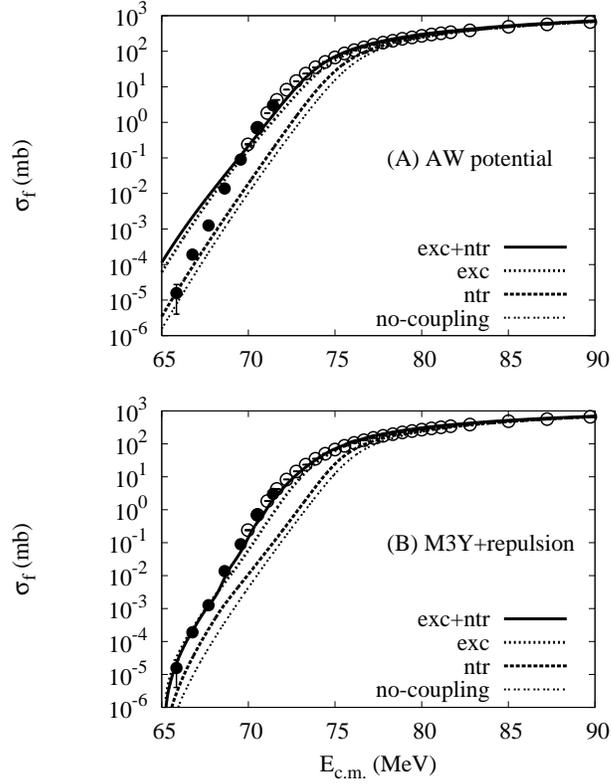}
\caption{\label{fusco} Calculated fusion cross sections, obtained with 
the AW potential (A)  and the M3Y+repulsion potential (B), 
are compared with the data of Ref. \cite{Morton2}
(open circles) and Ref. \cite{Morton3} (solid points). 
The curves show in increasing order the no-coupling limit, the
coupled-channels results for neutron transfer only (ntr) using 
$F_{1n}$ = 0.95, for surface excitations only (exc), and the combined 
effect of surface excitations and transfer (exc+ntr, solid curve).}
\end{figure}

\subsection{Fusion}

The measured fusion cross sections \cite{Morton2,Morton3} are compared
in Figs. \ref{fusco}A and \ref{fusco}B to the results of coupled-channels
calculations that are based on the AW and M3Y+repulsion potentials, 
respectively.
In each case we show in increasing order the fusion cross sections
we obtain in the no-coupling limit, by including
couplings to the one-neutron transfer (ntr), and to excitations 
of the surface modes (exc). The solid curves show the combined effect
of surface excitations and one-neutron transfer.
It is seen that the full calculation (solid curve) in Fig. \ref{fusco}B,
which is based on the M3Y+repulsion potential, provides the best fit to 
the data, in particular at the lowest energies where the 7 new data points 
\cite{Morton3} are shown by solid points. 

\begin{table}
\caption{Analysis of $^{16}$O+$^{208}$Pb fusion data. The best $\chi^2/N$, 
and the energy shift $\Delta E$ of the calculations that is required 
to minimized the $\chi^2$, are shown for two data sets.
The analysis included statistical errors and an assumed systematic 
error of 5\%.  The calculations were based on the AW potential and 
the M3Y+repulsion potentials, respectively, and included couplings 
to surface modes and one-neutron transfer using different values of 
the transfer coupling strength $F_{1n}$.}
\begin{tabular} {|c|c|cc|cc|}
\colrule
         &    & Old data & \cite{Morton2}    
& All data & \cite{Morton2,Morton3} \\
Potential & $F_{1n}$ & $\Delta E$ MeV & $\chi^2/N$ & $\Delta E$ MeV & 
$\chi^2/N$  \\
\colrule
AW  & 0    & -0.41  & 9.9 & 0.50 & 72 \\ 
    & 0.95 & -0.24  & 9.2 & 0.70 & 73 \\
    & 1.2  & -0.06  & 9.2 & 0.82 & 72 \\
\colrule
M3Y       & 0    & -0.61  & 14.2 & -0.45  & 25.6 \\
+         & 0.95 & -0.20  &  7.2 & -0.20  &  7.1 \\
repulsion & 1.10 & -0.08  &  5.7 & -0.10  &  5.6 \\
          & 1.20 &  0.0   &  5.2 &  0.015 &  5.2 \\
          & 1.30 &  0.1   &  5.8 &  0.10  &  5.4 \\
\colrule
\end{tabular}
\end{table}

The best $\chi^2$ per point we obtain by shifting the calculation
by an energy  $\Delta E$ is shown in Table II as a function of
the transfer strength $F_{1n}$. 
It is seen that the quality of the fit to the data is insensitive to the 
transfer strength when the calculations are based on the AW potential,
whereas the fit improves considerably with increasing transfer strength
when the M3Y+repulsion potential is used.
The best fit is achieved for $F_{1n}\approx$ 1.2.

It is also seen in Table II that the fit to all of the data points is very 
poor when the AW potential is used, whereas the $\chi^2/N$ is much smaller
when the calculations are based on the M3Y+repulsion potential.
We assumed in our analysis 
% that there is 
a 5\% systematic error
in addition to the statistical uncertainty.
However, it is not clear whether this assumption is realistic.
Another way of expressing the quality of the best fit is to say that it
requires a 12\% uncertainty, in addition to the statistical error, in
order to produce a $\chi^2/N \approx$ 1.

\begin{figure}
\includegraphics[width = 9cm]{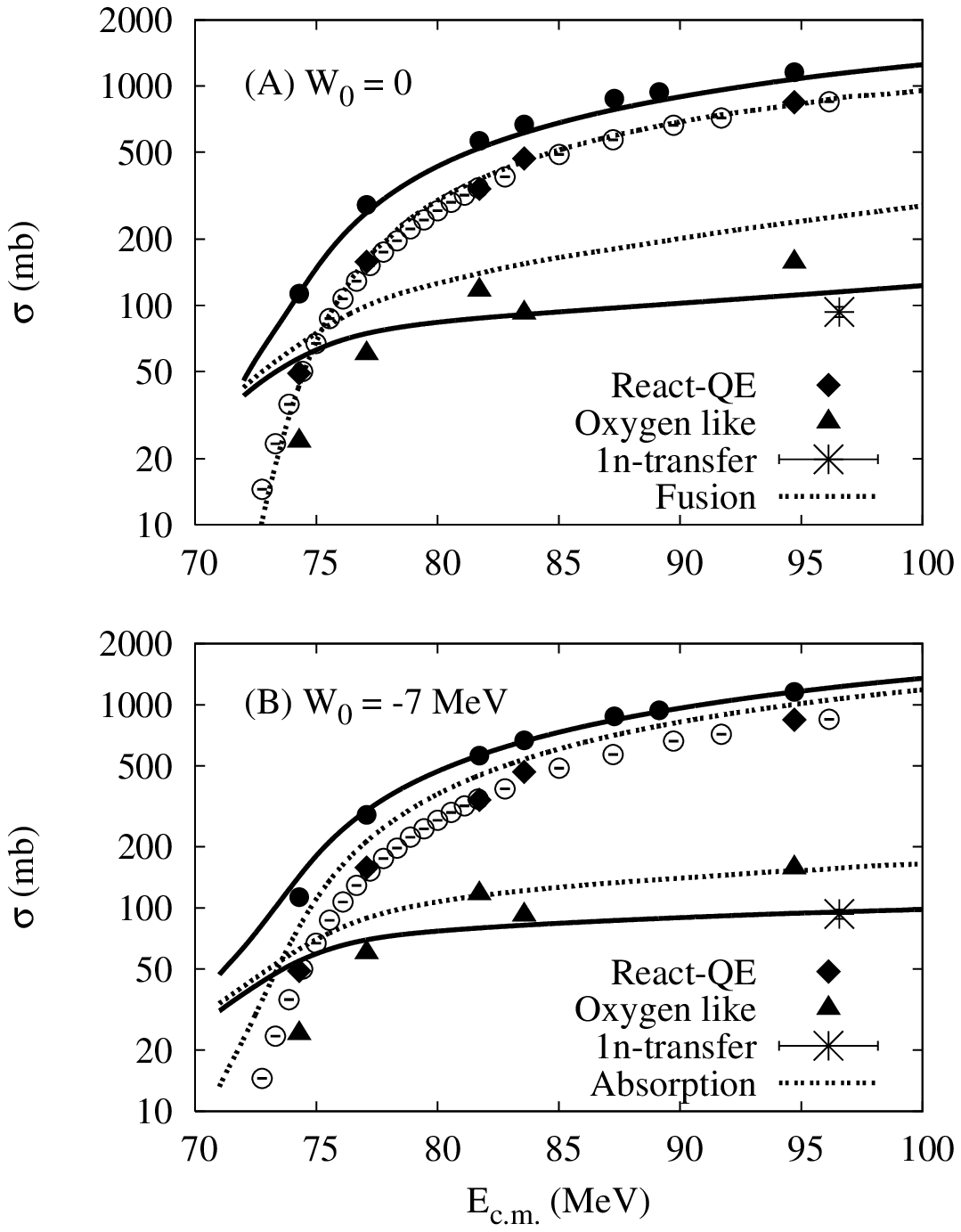}
\caption{\label{reacom3y} Calculated reaction cross sections (top solid 
curves) are compared to data (solid points) \cite{Flem}. 
The calculations are based on the M3Y+repulsion potential and include 
couplings to surface excitations and neutron transfer with $F_{1n}$ = 0.95.
The calculations in (A) use IWBC, whereas (B) employs an imaginary potential.
In decreasing order we also show the cross sections for fusion (open circles), 
oxygen like fragments (triangles), and one-neutron transfer (star).
The diamonds are the difference between the measured reaction and quasielastic 
cross sections.}
\end{figure}

\subsection{Reaction cross sections}

In this subsection we take a look at the measured reaction and
transfer cross sections \cite{Flem,Pieper} in order to determine a 
realistic value of the transfer coupling strength $F_{1n}$.
The reaction data are compared in Fig. \ref{reacom3y} to 
calculations that are based on the M3Y+repulsion potential
and include couplings to surface excitations and one-neutron
transfer with the coupling strength $F_{1n}$ = 0.95.
The results shown in Fig. \ref{reacom3y}A were obtained without 
using any imaginary potential, i.~e., the only absorption in 
this case is the fusion which is determined by the IWBC.
The top solid curve is the total reaction cross section which falls 
below the data \cite{Flem} (top solid points) so there is obviously
room for more reaction channels.  

The next set of data points in Fig. \ref{reacom3y}A are the fusion 
cross sections (open circles) which are reproduced fairly well by
the calculation (upper dashed curve). The diamonds show the 
difference between the measured total reaction and the quasielastic 
cross sections. They agree very well with the measured
fusion cross cross sections, so the total reaction cross section 
is essentially comprised of fusion and quasielastic scattering.

The triangles in Fig. \ref{reacom3y}A are the measured cross sections 
for oxygen like fragments, i.~e., the sum of the inelastic and neutron 
transfer cross sections.
These data points are slightly below the calculated values
(lower dashed curve). The lowest star-like point at 96.6 MeV
is the one-neutron cross section obtained in Ref. \cite{Pieper}
and it is also slightly below the calculated cross section
(the lower solid curve).

A simple way to simulate the reaction data in Fig. \ref{reacom3y} is to 
increase the transfer coupling strength $F_{1n}$. Thus we find that we 
need a value in the range of $F_{1n}\approx$ 1.2 - 1.3 in order to 
reproduce the total reaction cross section at the higher energies.
It is interesting that this coupling strength is roughly what 
produces the best fit to the fusion data according to Table II.
This implies that the calculated surface excitation plus one-neutron 
transfer cross section accounts in this case for the experimental 
quasielastic cross section.

Another way to account for the total reaction cross section is to employ 
a complex ion-ion potential.  We find that the total reaction cross section 
can be simulated quite well by including in the calculations an imaginary 
potential of the Woods-Saxon type with the parameters:
$W_0$ = -7 MeV, $R_w$ = 11 fm, and $a_w$ = 0.45 fm.
The results are shown in Fig. \ref{reacom3y}B.
The absorption cross section (upper dashed curve) exceeds the
measured fusion cross section because it must now 
simulate the sum of fusion and charged-particle transfer.
The calculated cross section for oxygen-like fragments (lower dashed curve)
is in good agreement with the data (triangles).
The measured one-neutron transfer cross section at 96.6 MeV \cite{Pieper} 
(the star-like symbol) is also reproduced by the calculation 
(the lower solid curve). This was achieved as mentioned earlier by 
adjusting the coupling strength to $F_{1n}$ = 0.95.

\begin{figure}
\includegraphics[width = 8cm]{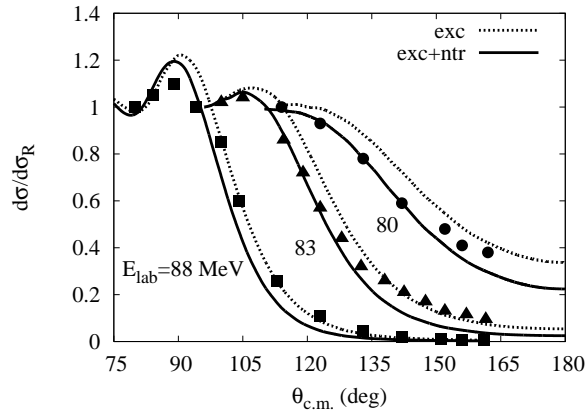}
\caption{\label{edelm3y} Elastic scattering data at $E_{lab}$ = 80, 83, 
and 88 MeV \cite{Flem} are compared with calculations that are based on 
the M3Y+repulsion and the imaginary potential discussed in the text. 
The dashed curves (exc) include the effect of surface excitations.
The solid curves (exc+ntr) include in addition the effect of 
neutron transfer using the strength $F_{1n}=0.95$.}
\end{figure}

\subsection{Elastic scattering}

The elastic scattering cross sections we obtain are compared 
with data \cite{Flem} in Fig. \ref{edelm3y}.
The calculations include the diagonal imaginary potential
($W_0$=-7 MeV, $a_w$=0.45 fm, $R_w$=11 fm) which was calibrated in 
the previous subsection so that the total reaction cross section
was reproduced. The dashed curves are based on couplings
to surface excitations (exc). The solid curves include in addition
the effect of transfer (exc+ntr) and they are seen to reproduce the 
data at the lowest energies.  Some discrepancies develop at the 
highest energy where the calculated rainbow peak is too high and 
the large angle scattering cross section is too low.
This is somewhat disappointing because the imaginary potential and the 
transfer strength $F_{1n}$ were adjusted in the previous subsection 
to account for the measured cross sections, c.~f. Fig. \ref{reacom3y}B.   

% NOT SHOWN: We can also see that the couplings to the excited states
% in \o16 play an important role in the calculations.
% This can be seen by comparing the `No $^{16}$O exc' 
% calculation with the exc-ntr calculation in Fig. \ref{edelm3y}.
% The two curves are shifted in angle, which to some extent
% is equivalent to a shift in beam energy.
% This may reflect the adiabatic renormalization effect
% discussed in Ref. \cite{Hagino97}. 

% Hagino et al. \cite{Hagino97} have pointed out that couplings to 
% the $3^-$ state in $^{16}$O provides an important polarization effect 
% and adiabatic renormalization which, however, does not affect 
% the structure of the calculated barrier distribution for fusion.  
% It is interesting to note that this conclusion does not hold
% when we consider the fusion at extreme sub-barrier energies.

\subsection{Barrier distribution and $S$ factor}

A good way to focus on the energy dependence of the fusion cross 
section at energies close to the Coulomb barrier is to plot the 
so-called barrier distribution, which is defined as the second 
derivative of the energy weighted fusion cross section \cite{Rowley}
\beq
B(E_{c.m.}) = \frac{d^2(E_{c.m.}\sigma_f(E_{c.m.}))}{dE_{c.m.}^2}.
\eeq
\begin{figure}
\includegraphics[width = 8cm]{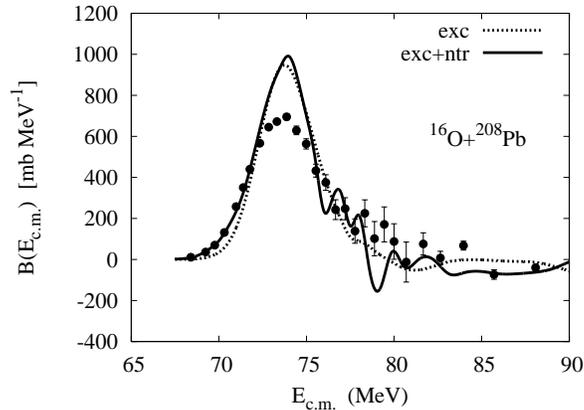}
\caption{\label{2dm3y} Barrier distributions extracted from the data
and from calculations that are based on the M3Y+repulsion potential.
Both calculations include couplings to surface excitations (exc). 
The solid curve (exc+ntr) includes in addition the coupling to neutron 
transfer with the coupling strength $F_{1n}$=1.2.}
\end{figure}
The results we obtain, with and without the effect of transfer, are
compared in Fig. \ref{2dm3y} to the barrier distribution we have
extracted from the data.  The most obvious discrepancy with the data 
is the much higher peak of the calculated distributions. 
The same problem was recognized in the coupled-channels calculations
presented in Ref. \cite{Morton2}. 
There the discrepancy was removed by applying a very small diffuseness 
($a\approx$ 0.4 fm) of the ion-ion potential but that would be 
inconsistent with the high-energy behavior of the fusion cross section 
which required a large diffuseness ($a\approx$ 1 fm) \cite{Morton2}.
It is unfortunate that using the shallow entrance channel potential 
we obtain with the M3Y+repulsion double-folding potential does not 
resolve the discrepancy with the peak height of the experimental 
barrier distribution.

\begin{figure}
\includegraphics[width = 8cm]{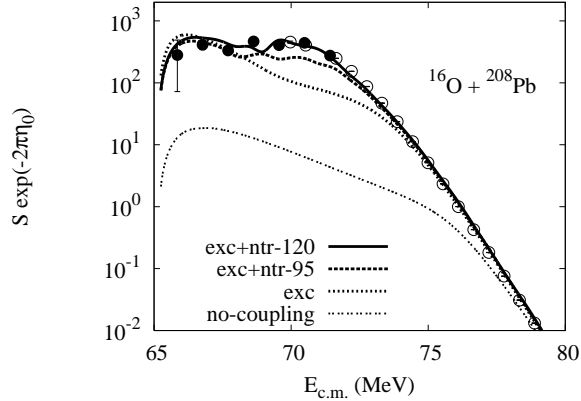}
\caption{\label{sfus} Calculated S factors for fusion (scaled by the factor 
$\exp(-2\pi\eta_0)$, where the reference Sommerfeld parameter $\eta_0$ is calculated 
at 75 MeV) are compared to the low-energy data \cite{Morton2,Morton3}. 
The curves show in increasing order the no-coupling limit, and coupled-channels 
results obtained with different transfer strengths,
namely, $F_{1n}$ = 0 (exc), 0.95 (exc+ntr-95), and 1.2 (exc+ntr-120).}
\end{figure}

A significant difference between the two calculations shown in 
Fig. \ref{2dm3y} is the behavior at the lowest energies. 
The calculation which includes the effect of transfer reproduces
the low energy data very well, whereas the calculation which is 
based on couplings to excited states only falls off too steeply.

A good way to emphasize the low-energy behavior of the fusion cross 
section is to plot the $S$ factor \cite{Jiangsys} for fusion,
\beq
S = E_{c.m.} \ \sigma_f(E_{c.m.}) \ \exp(2\pi\eta),
\eeq
where $\eta$ = $Z_1Z_2e^2/(\hbar v_{rel})$ is the Sommerfeld parameter,
and $v_{rel}$ is the asymptotic relative velocity in the entrance channel
of the reacting nuclei. The $S$ factors we obtain are shown in Fig. \ref{sfus}. 
It is seen that the calculation `exc', 
which includes couplings to surface excitations, makes a very poor 
fit to the data around 70 MeV. The calculation `exc+ntr-120', which  
in addition to surface excitations includes couplings to one-neutron 
transfer with the strength $F_{1n}$ = 1.2, makes a surprisingly good fit.
It is evident that couplings to transfer channels, in combination 
with the M3Y+repulsion potential, play a crucial role in explaining 
the fusion data at the lowest energies.
Calculations that are based on the AW potential, on the other hand,
do a very poor job at low energies (see Fig. \ref{fusco}A), and the 
quality of the fit to the data shows very little sensitivity to the 
transfer strength according to Table II.

\begin{figure}
\includegraphics[width = 9cm]{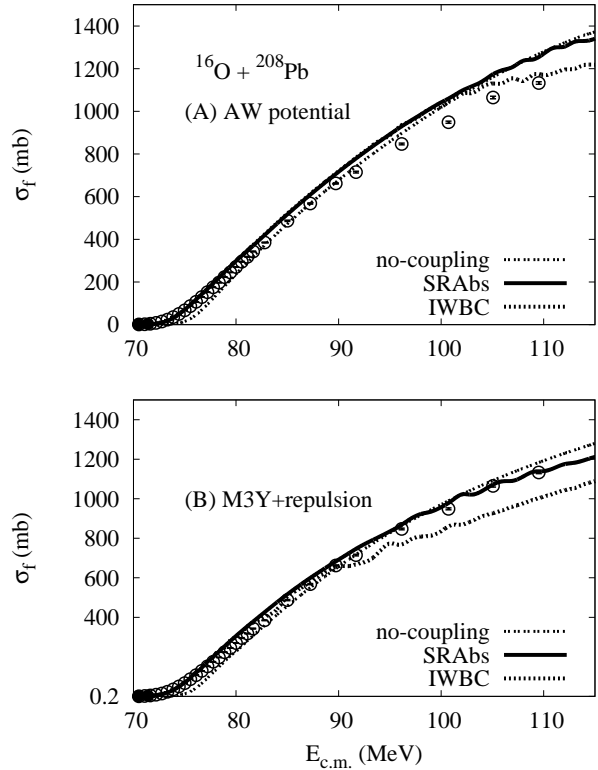}
\caption{\label{lfus} Linear plot of fusion cross sections
obtained with the AW potential (A) and the M3Y+repulsion potential (B).
The no-coupling limits (thin dashed curves) are compared to data 
\cite{Morton2} and to coupled-channels calculations 
(exc+ntr, using $F_{1n}$ = 0.95). The fusion obtained form IWBC 
(dashed curves) is supplemented with the absorption from a 
short-ranged imaginary potential (SRAbs, solid curves).}
\end{figure}

\subsection{Fusion at high energies}

The behavior of fusion at high energies is illustrated in Fig. 
\ref{lfus} which shows a linear plot of the cross sections obtained 
using the AW potential (A) and the M3Y+repulsion potential (B). 
The no-coupling limit which is based on the AW potential  
(Fig. \ref{lfus}A) is seen to exceed the data at the highest energies.
This is consistent with the analysis in Ref. \cite{Newton}
which showed that the fusion data are suppressed at high energy. 
This problem was fixed as mentioned earlier by using a large 
diffuseness of the ion-ion potential \cite{Morton2,Newton}.

The no-coupling limit which is based on the M3Y+repulsion potential 
is shown in Fig. \ref{lfus}B. It is in much better agreement with the 
data at the highest energies. Thus the application of the M3Y+repulsion 
potential seems to help resolve the problem of the suppression of 
the high-energy fusion data.
There is, however, another problem at high energy, namely, that 
the fusion cross sections obtained in coupled-channels calculations
from the IWBC tend to drop far below the no-coupling limit and even 
below the data.  This trend is clearly seen in Fig. \ref{lfus}B.

We have chosen to fix the problem with the IWBC in coupled-channels
calculations at high energies by supplementing the ion-ion potential 
with a short-ranged imaginary potential that acts near the location 
$R_{\rm pocket}$ of the minimum of the pocket in the entrance channel
potential.
(We use the Woods-Saxon parameters: 
$W_0$= -10 MeV, $a_w$ = 0.1 fm, $R_w$ = $R_{\rm pocket}$.)
It is seen that this prescription produces a cross section 
(solid curves in Fig.  \ref{lfus}) that is closer to the 
no-coupling limit.
Moreover, the agreement with the data is very good when we apply
the M3Y+repulsion potential (Fig. \ref{lfus}B), whereas the data 
are suppressed when compared to the calculations that are based
on the AW potential (Fig. \ref{lfus}A).

It is unfortunate that we have to resort to a short-ranged
imaginary potential in order to be able to reproduce the 
high-energy fusion data because this prescription does not 
work at extreme subbarrier energies.
We demonstrated that in section VII.A of Ref. \cite{Misicu2}
and it is also true for the $^{16}$O+$^{208}$Pb fusion reactions.
At the moment we are only able to reproduce the hindrance of
fusion at extreme subbarrier energies when we use IWBC
without any imaginary potential.

Evidently, the behavior of the high energy fusion cross section 
is not trivial. That may not be so surprising because it is also 
difficult to reproduce the total reaction cross section and the 
elastic scattering at high energy, without making resort to some 
kind of absorption.  
Another problem is that the rotating frame approximation which 
we have used is unrealistic at large angular momenta because it 
ignores completely the angular momentum dissipation which together 
with the energy dissipation can be critical for high energy fusion.  
The qualitative influence of angular momentum dissipation is 
nicely illustrated in Fig. 18 of Ref. \cite{Bock}.

\begin{figure}
\includegraphics[width = 8cm]{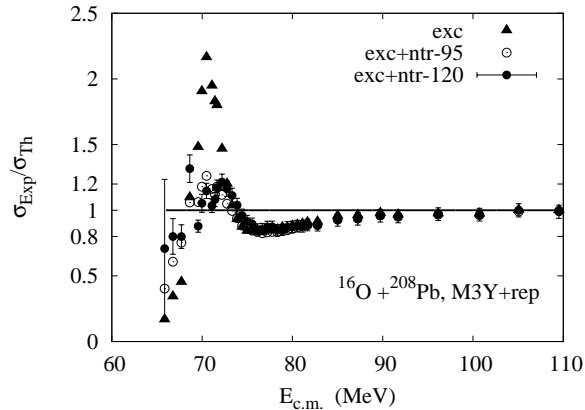}
\caption{\label{rm3y} Ratio of the measured and calculated fusion cross 
sections. The calculations are based on the M3Y+repulsion potential.
They include couplings to surface excitations (exc), and to surface 
excitations and one-neutron transfer using the coupling strengths 
$F_{1n}$ = 0.95 (exc+ntr-95) and 1.2 (exc+ntr-120).}
\end{figure}

The quality of the fits to the data is illustrated in Fig. \ref{rm3y} 
in terms of the ratio of the measured and calculated fusion cross sections.
The three coupled-channels calculations were all based on the M3Y+repulsion 
potential and they have  been shifted by the energy $\Delta E$ 
which is given in Table II, in order to produce the best $\chi^2/N$.
This requirement tends to force the calculations to be in good agreement 
with the high energy data, because the statistical error is very small
at high energy.  It is only when we use the AW potential that we see 
a suppression of the data at high energy (c.f. Fig. \ref{lfus}A).
Thus we conclude that using the M3Y+repulsion potential not only
explains the fusion hindrance phenomenon at low energies but it also 
helps eliminate the suppression of fusion that was observed in 
Ref. \cite{Newton} at high energies.

The main discrepancy between experimental and calculated cross sections
occurs in Fig. \ref{rm3y} at energies slightly below and slightly above 
the Coulomb barrier, which is located at $V_{CB}$ = 75.6 MeV.
The enhancement of the ratio $\sigma_{\rm exp}/\sigma_{\rm cal}$ just 
below the barrier is very sensitive to which calculation we compare to.
The reduction of the ratio just above the barrier, on the other hand, 
is rather insensitive to the calculation we consider.

It is not clear how one can eliminate the deviation of the cross section 
ratio from unity.  While the enhancement in Fig. \ref{rm3y} below the 
Coulomb barrier can be reduced by various 
means (by changing the coupling strengths, the number of channels, or 
by adjusting the ion-ion potential) the suppression above the barrier 
seems to be more robust. We note that the suppression above the barrier
is closely related to the large peak height of the calculated barrier 
distributions shown in Fig. \ref{2dm3y}.

\section{Conclusion}

We have shown that the hindrance of fusion, which has recently been 
observed in $^{16}$O+$^{208}$Pb reactions at low energies, is consistent 
with the shallow pocket in the entrance channel potential which is produced 
by the M3Y+repulsion double-folding potential.
There is strong evidence that couplings to transfer channels 
play a crucial role in explaining the energy dependence of the
fusion cross section (or $S$ factor) at the lowest energies.

The influence of couplings to transfer reactions in coupled-channels 
calculations of the $^{16}$O+$^{208}$Pb fusion has been recognized before
\cite{Steve,Ian1,Ian2} but the importance is much more dramatic
when the new low-energy fusion data \cite{Morton2} are considered 
and the coupled-channels calculations are based on the shallow 
M3Y+repulsion potential.
The best agreement with the fusion data is achieved by boosting
the neutron transfer coupling strength so that the calculations 
reproduce the measured total reaction cross sections. This is a 
nice consistency check of the coupled-channels calculations.

Another way to account for the observed reaction cross sections 
is to employ an imaginary potential, and this made it possible
for us to reproduce the elastic scattering data, at least at 
energies close to the Coulomb barrier.
There are still some problems in accounting for the scattering 
data at energies far above the Coulomb barrier, and the fusion,
which we usually define in terms of in-going wave boundary conditions,
has to be supplemented with the absorption in a short-ranged imaginary 
potential at high energies in order to be able to simulate the data.
Using this prescription at high energies, we are able reproduce 
the fusion data over eight orders of magnitude, from 10 nb to 1 b, 
with an average (root-mean-square) deviation of the order of 12\%. 
It is a challenge to theory to reduce this deviation further.

Since the coupling to transfer plays such a prominent role in the
fusion of $^{16}$O+$^{208}$Pb at very low energies, it may be useful in 
future work to reexamine the transfer form factors we have used.
They were developed for peripheral reactions (much the same way the 
Aky\"uz-Winther potential was developed to describe the elastic scattering 
in peripheral collisions) but they may not be realistic at shorter distances 
between the reacting nuclei.

{\bf Acknowledgments}

We are grateful to B. B. Back, M. Dasgupta, and C. L. Jiang for discussions
and encouragement.
This work was supported (H.E.) by the U.S. Department of Energy,
Office of Nuclear Physics, under Contract No. DE-AC02-06CH11357.

\section{Appendix: Density parametrization}

The matter or charge density of nuclei is often parametrized as 
$\rho_0 f((r-R)/a)$ in terms of the Fermi function
$f(x) = 1/(1+\exp(x)).$ For analytic purposes it is convenient to use 
instead the symmetrized form
$$\rho(r) = \rho_0 \ f((r-R)/a) \cdot f(-(r+R)/a)$$ 
\beq
\label{symff}
= \frac{{1\over 2} \rho_0 \exp(R/a)}{\cosh(r/a)+\cosh(R/a)}.
\eeq
The radial shape is essentially the same as that of the normal Fermi 
function, when $R$ is much greater than $a$. The largest modification
is at $r=0$, where the ordinary Fermi function is multiplied by the
factor $1/(1+\exp(-R/a))$.

A useful integral in this connection is
$$I(k) = \int_0^\infty dr \ \cos(kr) \ \rho(r)$$
\beq
\label{1Dk}
= 
{\rho_0 \exp(R/a)\over 2\sinh(R/a)} \ {a\pi \sin(kR)\over \sinh(ka\pi)},
\eeq
which follows from Eq. 3.983 no. 1 or no. 2 in Ref. \cite{GR}. 
One can also invert this expression
\beq
\label{1Dr}
\rho(r) = {2\over \pi} \int_0^\infty dk \ \cos(kr) \ I(k).
\eeq 
These are the one-dimensional Fourier transform relations between
$\rho(r)$ and $I(k)$. 

The three-dimensional Fourier transform of $\rho(r)$ can easily be 
derived from Eq. \ref{1Dk},
$$\rho({\bf k}) = \int d{\bf r} \ \exp(-i{\bf kr}) \ \rho(r)$$
\beq
\label{3Dka}
= \frac{4\pi}{k} \int_0^\infty dr \ r \ \sin(kr) \ \rho(r) 
= - {4\pi\over k} {d I(k)\over dk}.
\eeq
Inserting the derivative of (\ref{1Dk}) we obtain
$$\rho({\bf k}) = \frac{4\pi\rho_0 \exp(R/a)}{2 \sinh(R/a)} \ 
\frac{a\pi}{k} \times$$
\beq
\label{3Dkb}
{a\pi \sin(kR) \cosh(ka\pi) - R \cos(kR) \sinh(ka\pi)\over
(\sinh(ka\pi))^2}.
\eeq
A similar expression was derived in Ref. \cite{uberall},
Eq. (3-8j). It differs from Eq. (\ref{3Dkb}) by the factor
$\exp(R/a)/[2\sinh(R/a)]$, which is usually close to one.
The trick in deriving the analytic expression, Eq. (\ref{3Dkb}), was to 
use the symmetrized density distribution, Eq. (\ref{symff}).

The overall normalization of the density in terms of the mass number $A$
can be determined from Eq. (\ref{3Dkb}) evaluated at ${\bf k}={\bf 0}$,
\beq
\label{norm}
A = \rho({\bf k}={\bf 0}) =
\frac{4\pi\rho_0 \exp(R/a)}{2 \sinh(R/a)} \ 
\frac{R^3}{3} \ 
(1+\Bigr(\frac{\pi a}{R}\Bigr)^2).
\eeq
% This means that the normalization of the density in Eq. (\ref{symff}) is
% \beq
% \frac{1}{2} \ \rho_0 e^{R/a} =
% \frac{3A}{4\pi} \ \frac{sh(R/a)}{R(R^2+(a\pi)^2)}.
% \eeq
The mean square radius of the ground state density can be extracted 
from the $k^2$ term in the expansion of $\rho({\bf k})$,
$$\rho({\bf k})=A(1-{1\over 6} k^2<r^2>+....).$$
This yields the familiar result, Eq. (2-65) of Ref. \cite{BM1},
\beq
\label{rmsr} 
<r^2> = {3\over 5} \ (R^2 \ + \ {7\over 3} \ (a\pi)^2).
\eeq

\end{document}